# High Electron Mobility and Large Magnetoresistance in the Half-Heusler Semimetal LuPtBi

Zhipeng Hou[1,2], Wenhong Wang[1,*], Guizhou Xu[1], Xiaoming Zhang[1], Zhiyang Wei[1], Shipeng Shen[1], Enke Liu[1], Yuan Yao[1], Yisheng Chai[1], Young Sun[1], Xuekui Xi[1], Wenquan Wang[2], Zhongyuan Liu[3], Guangheng Wu[1] and Xi-xiang Zhang[4]

[1]State Key Laboratory for Magnetism, Beijing National Laboratory for Condensed Matter Physics, Institute of Physics, Chinese Academy of Sciences, Beijing 100190, China

[2]College of Physics, Jilin University, Changchun 130023, China

[3]State Key Laboratory of Metastable Material Sciences and Technology, Yanshan University, Qinhuangdao 066004, China

[4]Physical Science and Engineering, King Abdullah University of Science and Technology (KAUST), Thuwal 23955-6900, Saudi Arabia.

## Abstract

Materials with high carrier mobility showing large magnetoresistance (MR) have recently received much attention because of potential applications in future high-performance magneto-electric devices. Here, we report on the discovery of an electron-hole-compensated half-Heusler semimetal LuPtBi that exhibits an extremely high electron mobility of up to 79000 $cm^2/Vs$ with a non-saturating positive MR as large as 3200% at 2 K. Remarkably, the mobility at 300 K is found to exceed 10500 $cm^2/Vs$, which is among the highest values reported in three-dimensional bulk materials thus far. The clean Shubnikov-de Haas quantum oscillation observed at low temperatures and the first-principles calculations together indicate that the high electron mobility is due to a rather small effective carrier mass caused by the distinctive band structure of the crystal. Our finding provide a new approach for finding large, high-mobility MR materials by designing an appropriate Fermi surface topology starting from simple electron-hole-compensated semimetals.

[*] Corresponding author. wenhong.wang@iphy.ac.cn

## I. INTRODUCTION

The search for new materials with both high charge-carrier mobility and magnetoresistance (MR) is an important topic of research in condensed matter physics and for various device applications. In contrast to the large negative MR often observed in magnetic materials,[1-6] several non-magnetic material systems, ranging from narrow gap semiconductors[7-9] and zero-gap semimetals [10-13] to two-dimensional transition metal oxides[14] and topological insulators,[15,16] have recently been discovered to show a large non-saturating positive MR. Attempts to explain this high MR include a classical MR effect based on a long transport mean free path due to mobility distribution[17] and the quantum effect with a linearly dispersing band structure.[18] Because materials with high MR and carrier mobility show promising applications for future high-performance spintronics devices, more investigation into the mechanism is warranted.

Compensated semimetals have approximately the same number of holes as electrons (a balanced hole-electron resonance condition) and thus, a zero band-gap at the Fermi surface. For these reasons, they are expected to be an ideal platform for exploiting a high positive MR; well-known examples include Bi and graphite.[10,11] Recently, the compensated semimetal $WTe_2$ was shown to have a non-saturating positive MR of up to 450,000% at low temperatures,[19] which would make it suitable for investigating the unusual transport properties and carrier scattering mechanisms of compensated semimetals. However, layered $WTe_2$ loses its advantage at higher temperatures where its positive MR reduces significantly due to the dramatic change in the Fermi surface, making it impractical for use in spintronics devices because they operate at a wide temperature range.[20]

In parallel, many ternary half-Heusler compounds with 18 valence electrons that are naturally semimetals with tunable structural and electronic properties have been well studied for their multifunctional properties and their potential application as thermoelectric materials. [21,22] Recently, studies have focused on some heavy half-Heusler compounds composed of the rare earth elements Ln (Ln = Sc, Y, La, Lu) such as LnPtSb and LnPtBi. These compounds are of interest because they feature a zero-gap topologically non-trivial band inversion in the presence of spin-orbit coupling and may exhibit unprecedented features, such as high carrier mobility and MR when subjected to a magnetic field $H$.[23-28] Despite considerable efforts, the highest carrier mobility observed in half-Heusler semimetals, like LaPtBi and YPtBi, is limited to ~ 4000 cm$^2$/Vs at room temperature (RT).[26] Here, in this paper, we report an electron-hole-compensated half-Heusler semimetal LuPtBi with an electron mobility as high as 79000 cm$^2$/Vs at 2 K. Remarkably, the mobility at 300 K is found to exceed 10500 cm$^2$/Vs, which is comparable to the 3D Dirac semimetal Cd$_3$As$_2$.[13,29] Note that LuPtBi crystal MR increases from 260 % at RT to 3200 % at 2 K. Combined ultrahigh mobility and MR across a wide temperature range makes compensated semimetal LuPtBi and related half-Heusler compounds potential material candidates for application to future spintronics devices.

## Ⅱ. EXPERIMENTAL METHODS AND STRUCTURE DETERMINATION

Single LuPtBi crystals were grown by a Bi-rich self-flux method. For the growth of LuPtBi crystals, the high-purity Lu (ingot, 99.99%), Pt (ingot, 99.99%) and Bi (ingot, 99.99%) starting materials were mixed together in a molar ratio of 1:1:10, and afterward the mixture was placed in an alumina crucible. This process was performed in an argon-filled glove box, where the oxygen and humidity content was less than 0.5 ppm. The whole assembly was sealed inside a tantalum tube under Ar gas followed by sealing in an evacuated quartz tube. Crystal growth was performed in a furnace by heating the tube from room temperature to 1150℃ over a period of 15h, maintained for 24h, and then slowly cooled to 650℃ at a rate of 2K/h. The excess Bi flux was removed by spinning the tube in a centrifuge at 650℃. After the centrifugation process, most of the flux contamination was removed from crystal surfaces and the remaining topical flux was etched by diluted hydrochloric acid. As shown in Fig. 1(c), the as-grown crystals were typically 1.5 × 1.2 × 0.8 mm$^3$ with mirror-like surfaces and were robust in air. As shown in Fig. 1 (d), the orientation of the as-grown crystals was first checked by X-ray diffraction and, the sharp (111) and (222) reflections indicate the crystal is oriented along the [111] axis. In addition, the inset of Fig. 1(d) shows the Laue diffraction pattern of single-crystal LuPtBi generated with the beam axis coincident with the [111] zone axis. Apparently well-developed (111) planes can be observed, suggesting that our single crystals are of high quality.

Composition of all well-polished single crystals was determined by Energy Dispersive X-ray (EDX) spectroscopy, equipped on a Hitachi S-4800 scanning electron microscope (SEM). EDX measurements were performed at different positions on crystal surfaces within an instrument error of 1-2%. EDX analyses were performed at numbered points of samples and the results are listed in Fig. S1 and Table S1, respectively, in the Supplemental Material.[30] No evidence of a Bi thin film or nanocluster was observed. According to element mapping, as shown in Fig. 2(a), Lu, Pt, and Bi are uniformly distributed across the surface, confirming the absence of a Bi thin film or cluster. The powder X-ray diffraction (XRD) measurements of pulverized crystals showed very sharp and strong peaks, which can all be indexed to a MgAgAs-type structure (*F-43m* space group) [See inset of Fig. 2 (b)]. The lattice constant was calculated as *a*=6.586 Å , which is in agreement with previous studies.[31]

As shown in Fig. 3(a), the cross-sectional cuts of a selected LuPtBi crystal for scanning transmission electron microscope (STEM) measurements were prepared using a dual-beam focused ion beam (FIB) along the [110] zone axis. An aberration-corrected STEM was performed using a JEOL 2100F (JEOL, Tokyo, Japan) transmission electron microscope operated at 200 keV. The microscope was equipped with a CEOS (CEOS, Heidelberg, Germany) probe aberration corrector. The attainable spatial resolution of this microscope is 90 picometers at an incident semiangle of 20 mrad. As shown in Figs. 3 (b) and (c), electron diffraction patterns and differently scaled STEM-ADF images HREM along the [100] stacking axis, taken from many fragments of LuPtBi, showed no sign of stacking faults, intergrowths or amorphous regions, indicating that crystal quality is good even at the nanoscale. None of the diffraction experiments indicated the presence of long- or short-range ordering of bismuth in crystals.

Electrical leads in the four-probe and Hall bar configurations were attached onto the samples using silver paste by platinum wires. Transport and specific heat measurements were performed in between 2 and 300 K on a Quantum Design physical properties measurement system. More than 20 LuPtBi crystals were selected and then polished to a near-rectangular shape with 1.5×0.5 mm$^2$ in the (111) plane and a roughly 0.2-mm thickness for magnetotransport measurements. Table S2 and S3 in the Supplemental Material lists the transport parameters measured in six representative samples.[30] A conventional four-probe method was used for both out-of-plane resistivity and Hall measurements from 2 to 300 K in a magnetic field up to 14 T with a commercial apparatus (Quantum Design, model PPMS). The Band structure calculations used the full-potential APW method implemented in the WIEN2K package including spin-orbit coupling. [32]

## Ⅲ. RESULTS AND DISCUSSIONS

### A. Temperature and magnetic field dependence of resistivity

Figure 4(a) shows the *T*-dependent resistivity $\rho_{xx}$ for the six representative samples measured at a zero field. Above 50 K, $\rho_{xx}$- *T* curves of all samples are similar. However, as *T* decreases from 50 to 2 K, the values of $\rho_{xx}$ for S1 and S3 decrease rapidly, implying a strong enhancement of metallicity and crystal quality as well. This enhancement can be further confirmed by examining the residual resistivity ratio RRR [RRR=$\rho_{xx}(300K)/\rho_{xx}(2K)$]. As shown in Fig. 4(a), the sample S1 had the highest value of RRR=4.2, indicating a very low residual resistivity $\rho_{xx}(2K)=3.7\mu\Omega$cm, while for the sample S12 with RRR=1.4 and therefore it has the highest residual resistivity $\rho_{xx}(2K)=78\mu\Omega$ cm. As shown in Fig. 4(b), from the MR data at 2K, we were able to determine that the residual resistivity and MR are closely related with the residual resistivity ratio (RRR). As the RRR increases, MR roughly increase, however, residual resistivity decreases. In general, the RRR value is a direct measure of defect concentration in single-crystal samples. Therefore, we can conclude that single-crystal quality is an important determinant of half-Heusler LuPtBi transport properties.

In Fig. 5(a), we show the *T*-dependent $\rho_{xx}$ for S1 measured at a series of applied magnetic fields. Interestingly, the applied fields change not only the *T*-dependent behavior but also significantly enhance its resistivity, especially at low temperatures. When the field is higher than 2 T, the material begins to demonstrate semiconducting-like resistivity behavior.[33] At 2K, residual resistivity increases more than 30 times (from 3.7 to 120 μΩcm) as the field increases from 0 to 10 T. We propose that the quality of crystals is the key for understanding the large sample to sample variation of resistivity.

Figure 5(b) shows MR measured at different *T* for S1 in magnetic fields up to 10 T. The MR value is defined as MR = [$\rho$(H)-$\rho$(0)] / $\rho$(0) × 100%, where $\rho$(*H*) and $\rho$(0) are the resistivities at field *H* and the zero field, respectively. An non-saturating positive MR, as large as 3200% was obtained at 2 K and *H*=10 T. When we rotate the magnetic field from the [111] to the [1-10] direction, the positive MR decreases to 2200% at *H*=10 T (See Fig. S2), indicating a much smaller transport anisotropy than in layered materials, such as $PdCoO_2$[8] and $WTe_2$.[9] We note that although at *H*=10 T in the [111] direction, the MR is significantly suppressed by thermal fluctuation with temperature increasing from 2 to 300 K, it is still a considerable 260%. This value is comparable with the compensated semimetal Bi [4, 34] but much larger than that of layered $PdCoO_2$[8] or $WTe_2$.[9]

Figure 5(c) presents the so-called Kohler plot [35], in which $\Delta\rho_{xx}(H)/\rho_{xx}(0)$ is plotted as a function of $\mu_0 H/\rho_{xx}(0)$. It is evident that the data obtained at high temperatures and low fields can be well fitted to the $H^{1.5}$-dependence expected for an orbital motion of conduction charge carriers instead of to Kohler's rule. This behavior suggests the coexistence of two kinds of carriers with different mobilities in the sample. On the other hand, all high field data measured at low-temperatures collapse onto a single curve that scales linearly with *H*, behavior described by Kohler's rule (See Fig. S3), which implies that at the low temperatures and high field only one type of carrier dominates the electrical transport properties in this compound.

**B. Hall resistivity and thermoelectric power**

To gain a further insight into the carrier transport in LuPtBi, we performed the Hall effect and thermoelectric power measurements. Fig. 6(a) shows the magnetic field dependence of the Hall resistivity $\rho_{xy}$ at different temperatures. We note that, at all the temperatures, the slope of $\rho_{xy}$ is negative at low fields, but it changes to positive in high-fields [see the enlarged figure in the inset of Fig. 6(a)]. The negative slope of $\rho_{xy}$ indicates the predominance of electron carrier contribution to the transport at low-fields, however, hole carrier takes over the role when field is large enough. By considering the highly nonlinear field dependence behavior of $\rho_{xy}$, e.g. negative slope at low-fields and sign change, we can thus propose that LuPtBi has two kinds of carriers with different temperature and field-dependent behaviors.

To confirm the above assumptions, we have further measured the temperature dependence of thermoelectric power ($S$) at $H = 0$ and 9 T, respectively as shown in Fig. 6 (b). The thermoelectric power data of LuPtBi are consistent, in a qualitative way, with the results obtained in the Hall measurements. On the one hand, at $H = 0$ T, the value of $S$ is negative at 300 K, signifying that the electron carrier dominates the thermoelectric transport. Upon cooling, the sign of $S$ shows a reversal at about 80 K, suggesting the transport properties are dominated by hole carriers at $T < 80$ K. On the other hand, at $H = 9$ T, the value of $S$ only shows a positive sign in the whole temperature range studied, which suggests a significant enhancement in the contribution of hole carriers at high-fields, agreeing well with the Hall data. Another remarkable result is that the temperature dependences of $S$ at $H = 0$ and 9 T are quite different, especially at $T > 80$ K. In contrary to a sign reversal, $S$ at 9 T exhibits a new broad peak at $T = 80$ K. Further discussion of the sign anomaly is given below (see Discussion section). We argue that the sign changes in Hall and $S$ signal at high-fields can be related to a field-induced depopulation of electron pockets and the possible Fermi surface rearrangement.

## C. Carrier mobility and concentration

In the next, we investigated a temperature-dependent role on conductivity tensors by the two-band model [33] to determine the mobility and concentration of electron and hole carriers, respectively. The longitudinal conductivity $\sigma_{xx}$ and Hall conductivity $\sigma_{xy}$ can be described as:

$$\sigma_{xx} = \frac{\rho_{xx}}{\rho_{xx}^2 + \rho_{xy}^2}, \qquad (1)$$

$$\sigma_{xy} = \frac{-\rho_{xy}}{\left(\rho_{xx}^2 + \rho_{xy}^2\right)}. \qquad (2)$$

Thus, in a low-field regime (0-1.2 T), carrier concentrations and mobilities can be independently extracted by fitting $\sigma_{xx}$ and $\sigma_{xy}$ with the two following Equations:

$$\sigma_{xx}(H) = \frac{n_e e \mu_e}{1 + \mu_e^2 H^2} + \frac{n_h e \mu_h}{1 + \mu_h^2 H^2} \qquad (3)$$

$$\sigma_{xy}(H) = \frac{n_e e \mu_e^2 H}{1 + \mu_e^2 H^2} + \frac{n_h e \mu_h^2 H}{1 + \mu_h^2 H^2} \qquad (4)$$

where $n_e$ ($n_h$) and $\mu_e$ ($\mu_h$) indicate the carrier concentrations and carrier mobilities of electrons (holes), respectively. The fitting parameters are dependent on temperature $T$ but independent from external magnetic field $H$. Using the fitting results both for Hall conductivity, $\sigma_{xy}$ [Fig. 7(a)], and longitudinal conductivity, $\sigma_{xx}$ [Fig. 7(b)], we could independently obtain the Hall and longitudinal mobilities and concentrations for electron and hole carriers, respectively. As shown in Figs. 7(c) and (d), both the Hall and longitudinal mobilities and carrier concentrations coincide well across the entire temperature range, indicating the validity of the two-carrier model as our fitting procedure. In addition, as shown in Fig. 7(c), in the whole temperature regime, hole carrier concentrations $n_h$ were slightly higher than those of electron carriers $n_e$, indicating that LuPtBi has a compensated nature. Notably, as shown in Fig. 7(d), electron mobility, $\mu_e$, is determined as 79000 cm$^2$/Vs at 2 K, which is comparable to that of the phonon glass semimetal $\beta$-CuAgSe [36]. More importantly, $\mu_e$ at 300 K is found to exceed 10500 cm$^2$/Vs. To date, this is the highest mobility reported at RT in any known half-Heusler semimetal and it is comparable with the compensated semimetal Bi [34] and Dirac semimetal Cd$_3$As$_2$.[29] In addition, the mobility of electron is always higher than that of hole, and both values were found closely related to the RRR (See Fig. S4). At 2 K, the electron-hole-compensated sample S1 had the highest mobility (79000 cm$^2$/Vs), MR (3200%) and RRR, versus the hole-dominated sample S12, which had the lowest RRR, mobility (1200 cm$^2$/Vs) and MR ratio (136%). Therefore, we suggest that the use of high-quality compensated semimetals is necessary to obtain the balance between electrons and holes that is required to achieve an ultrahigh MR, similar to that of WTe$_2$.[9]

In Fig. 8, we compared temperature dependence of carrier mobility normalized by $\mu$ (T = 2 K) for several of the most researched high-mobility systems. The mobility of SrTiO$_3$/LaAlO$_3$ [37] hetero-interface (blue) and La-doped SrTiO$_3$ [38] films (dark cyan) dropped rapidly with increasing temperature, while LuPtBi (red) exhibited only slight temperature dependence, preserving a comparatively high carrier mobility at room temperature. Such a weak temperature-dependent carrier mobility is comparable to the Dirac semimetal Cd$_3$As$_2$,[13] and can be quantitatively explained based on three distinctions of electron scattering: ionized impurity scattering (II), acoustic phonon scattering (AC) and optical phonon scattering (usually longitudinal, LO). [39] By applying the form used in Ref. 39 with $\mu_{AC} = \text{const.} \times \ln(1 + e^{E_F/k_B T})$, $\mu_{LO} = \text{const.} \times (e^{\hbar w_l/k_B T} - 1)$, we can reproduce the mobility changes (black line) of the Bi film (orange),[39] Ni-doped CuAgSe (green) [36] and LuPtBi (red) with varying parameters. As indicated by the dotted line, longitudinal optical phonon scattering is a primary source of carrier scattering in these materials at high temperatures. [40] Therefore, we assume that the weak dependence of mobility on temperature in LuPtBi can be attributed to the significant reduction in scattering from longitudinal optical phonons as a consequence of peculiarities in band structure.

**D. Shubnikov-de Hass Quantum oscillations.**

To understand the electronic states that contribute to transport properties, we investigate the MR and Shubnikov-de Hass (SdH) quantum oscillations at high fields. In Fig. S5, we show the temperature dependence of SdH quantum oscillations at high magnetic fields for samples S1 and S3. The field is applied parallel to the [111] direction. At high fields, the clear SdH oscillations are superimposed on a huge background of positive MR. In Fig. 9(a), we highlight the SdH oscillations by plotting $d^2\rho_{xx}/dH^2$ as a function of the inverse magnetic field ($1/H$) for sample S1. As seen in the inset, prominent SdH oscillations become visible at $H > 5$ T indicating the extreme mobility of carriers in this sample. We confirm that SdH oscillations are observable when $H$ changes from 0° to 90°, that is from [111] to [1-10] direction, suggesting their 3D origin (See Fig. S6). The details of the SdH oscillations in all geometries will be discussed elsewhere. The inset of Fig. 9 (a) shows the fast Fourier transform (FFT) spectra of oscillation at 0°, revealing two peaks at $F_1$=80 T and $F_2$=200 T, which are confirmed as the electron and hole Fermi pockets, respectively (See Fig. S7). By using the Onsager relationship, $F = (\hbar/2\pi e)A_F$, where $e$ is the electron charge and $\hbar$ is Planck's constant and $A_F$ is the cross-sectional area of the Fermi surface normal to the magnetic fields. The sizes of the electron and hole Fermi pockets are 0.01 Å$^{-2}$ and 0.019 Å$^{-2}$, respectively. The corresponding Fermi wave vector $\kappa_F$ for the electron and hole Fermi pockets are 0.048 Å$^{-2}$ and 0.086 Å$^{-2}$, respectively. According to the standard Lifshitz-Kosevich theory,[41] the cyclotron effective mass of the carriers ($m*$) can be obtained by fitting the temperature dependence of the normalized FFT amplitudes with a thermal damping factor, $R_T = \dfrac{2\pi^2\kappa_B T/\hbar\omega_c}{\sinh[2\pi^2\kappa_B T/\hbar\omega_c]}$, where $\kappa_B$ is Boltzmann's constant, $T$ is temperature and $\omega_c = eH/m*$ is the cyclotron frequency, which directly results in the effective mass $m*$. As shown in Fig. 9 (b), the effective masses for the electron and hole Fermi pockets yielded by the fits are $0.11\,m_e$ and $0.23\,m_e$ ($m_e$ is the free electron mass), respectively. Finally, the Fermi velocity for the electron and hole pockets are calculated using $v_F = \hbar\kappa_F/m^*$ are $5.5\times10^7$ cms$^{-1}$ and $3.3\times10^7$ cms$^{-1}$, respectively. Both values are close to that of the Dirac semimetal Cd$_3$As$_2$ [29, 42] and Weyle semimetals TaAs [43] and NbP [44]. Thus, the large Fermi velocity and small effective mass are responsible for the observed ultrahigh mobility in LuPtBi. Note that although the quantum limit is actually not reached in our LuPtBi crystal, there is clearly more than one Landau level occupied in our field range. Therefore, the appearance of a linear MR without reaching the quantum limit requires further investigation.

**E. Band structure and Fermi surface**

Band structure calculations provide insight into the origin of an electronic structure with ultrahigh mobility and MR such as in LuPtBi. As shown in Fig. 10 (a), the profile of density of states (DOS) reveals that the Fermi level (E$_F$) of LuPtBi is located at the valley of the DOS, confirming that the states at the Fermi level are dominated by the states of Pt and Bi atoms. Our band structure calculations are illustrated in Fig. 10 (b), and similar to a previous report, [45] valence and conduction bands barely cross the E$_F$ and do so at different places, indicating that LuPtBi is an electron-hole-compensated semimetal. This can also be confirmed by the volume of Fermi electron and hole pockets showing in Fig. 10 (c), which implies Fermi surface compensation in LuPtBi. The Fermi surface of LuPtBi consists of eight equivalent needle-like electron pockets and one hole pocket located almost isotropically at the center of the Brillouin zone along crystallographic [111] directions; for each crystallographic [111] direction, two needle-like electron pockets locate symmetrically with the Γ point, and the angle of two adjacent ellipsoid pockets are 70.52°. At the crystallographic (111) plane, six equivalent needle-like electron pockets locate symmetrically around the [111] axis and each electron pocket can be obtained by a 60° rotation around the [111] axis. This exceptional feature of the electronic structure is reminiscent of that of the well-known semimetal Bi [10, 33] and layered alloy NbSb$_2$.[46] The unique shape of the Fermi surface of LuPtBi crystals is energetically unfavorable; therefore, a magnetic field can cause the contacting *p*-symmetry band to split such that a bandgap appears at the Fermi level, which may cause the carrier concentration and the mobility tensor change with magnetic field. On the other hand, the highly anisotropic electron pockets may cause large variations in mobility, which may also help to explain its unusual large positive MR. In this regard, further angle-resolved photoemission spectroscopy experiments are necessary to verify the Fermi surface topology and its evolution with *H* from which the large non-saturating linear MR may originate.

The next question to be addressed is why the signs of the Hall resistivity and thermoelectric power data change in high-fields. In fact, the sign change of the Hall resistivity with *H* only occurs when the two kinds of carriers have distinct mobility and/or concentration. In our LuPtBi sample, the high-mobility electrons with a small effective mass dominate the electrical transport and gives rise to the negative Hall resistivity in low-fields. In order to achieve the sign reversal of Hall resistivity and thermoelectric power in high-fields, the number of high-mobility electrons should be much smaller than the number of low-mobility holes. In this case, the sign change is reminiscent of that of the change of the Fermi surface topology (field-induced Lifshitz transition[47]) in momentum space in the gapless materials, such as HgTe [48] and half-Heusler CePtBi. [49] The scenario is that the small Fermi surface is energetically unfavorable in high-fields. Therefore, the magnetic field may cause the displacement of the energy-band edges, which may further induce the contacting *p*-symmetry band split, and as a result, the concentration of high-mobility electrons will decrease while the concentration of low-mobility hole will increase instead. Further study will be required to verify this.

## IV. CONCLUSION

We end by summarizing the important features of half-Heusler LuPtBi semimetal. First, it shows an extremely high electron mobility up to 79,000 cm$^2$/Vs together with a non-saturating positive MR as large as 3200% and at 2 K. Remarkably, the electron mobility at 300 K is found to exceed 10,500 cm$^2$/Vs, which is among the highest values reported in 3D bulk materials thus far. Second, it has a distinctive compensated band structure showing a rather small effective carrier mass as derived from the SdH quantum oscillations. The small effective carrier mass is responsible for the extremely high electron mobility and large positive MR. Third, its high-mobility electron conduction is robust against thermal fluctuation, as evidenced by the weak temperature-dependent electron mobility that results from a significant reduction in scattering from longitudinal optical phonons. Last, the electron-hole-compensated band structure is rather sensitively to the applied magnetic fields, as evidence by the observation of the signs of the Hall resistivity and thermoelectric power in high-fields. However, it is remains to be investigated how and why the fields can influence the Fermi surface topology. Nevertheless, our results indicate that, even in the reported 3D semimetals, both the ultrahigh mobility and large positive MR may be obtained by the appropriate design of their band structure with the use of the Fermi surface compensated strategy. We expect that the strategy would provide useful guidelines for the development of ultrahigh carrier mobility semimetal showing large positive MR as high-performance magnetoresitive devices in other known half-Heusler semimetals and open an area of research of both fundamental and applied importance.

**ACKNOWLEDGMENTS**

We thank Prof. J. R. Sun and Prof. C. Felser for fruitful discussions. This work is supported by the National Basic Research Program of China (973 Program 2012CB619405) and the National Natural Science Foundation of China (Grant Nos. 51171207and 11474343), and Strategic Priority Research Program B of the Chinese Academy of Sciences under the grant No. XDB07010300.

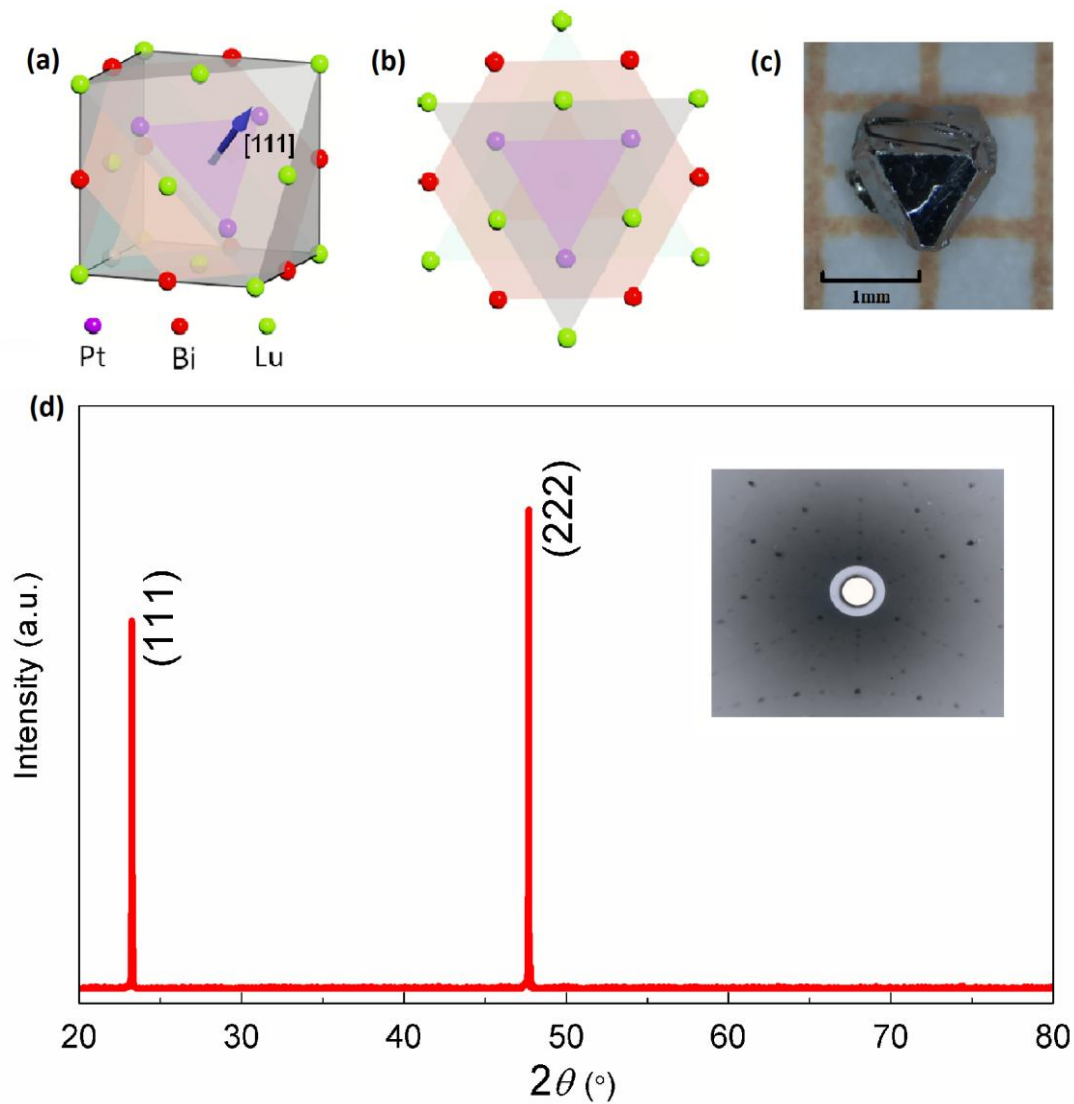

**FIG. 1.** (color online) (a) The crystal structure of LuPtBi. The blue arrow shows the normal of natural cleavage plane along the (111) surface. (b) The projected view of lattice along the [111] direction, displaying Lu, Pt and Bi stacked layers. (c) Photographs of synthesized single-crystal LuPtBi placed on a millimeter grid. One division represents 1 mm in the diagram; crystals were grown to $1.5 \times 1.0 \times 0.8$ mm$^3$, with mirror-like surfaces and were robust in air. (d) XRD patterns with the x-ray along perpendicular direction of hexagonal surfaces are presented.

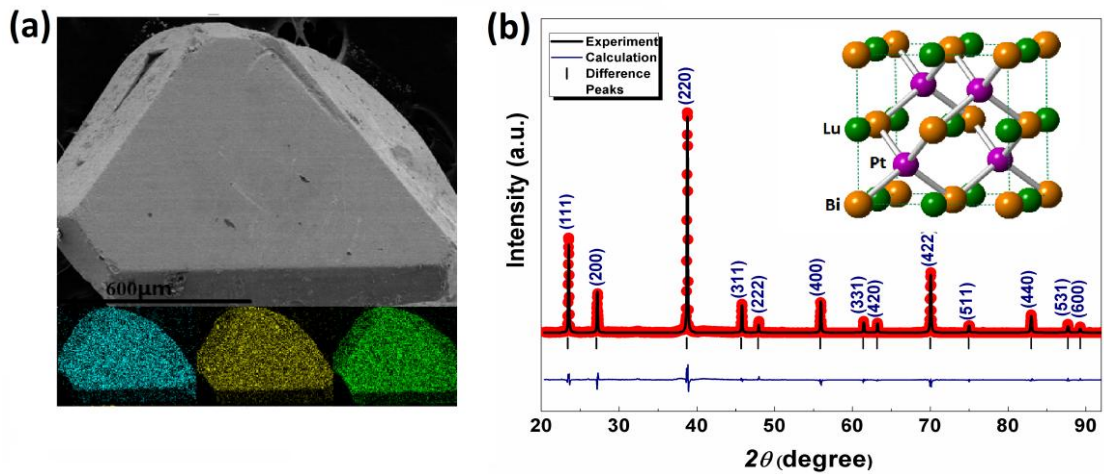

**FIG. 2.** (color online) (a) A SEM image of a typical single crystal of LuPtBi after removing excess Bi flux; bottom plane. The color maps are elemental distributions of Lu (bule), Pt (yellow) and Bi (green) acquired by scanning EDX. (b) Observed (red line) powder XRD patterns of crushed LuPtBi single crystals at room temperature show the results of structural refinement (green circles). The pink segments indicate the expected diffraction peaks. LuPtBi reflections are indexed within the MgAgAs-type structure (space group *F43m*, 216) and the refined lattice parameter is $a = 6.5861$ Å. The inset shows a structural view of conventional LuPtBi unit cell, which four formulary units.

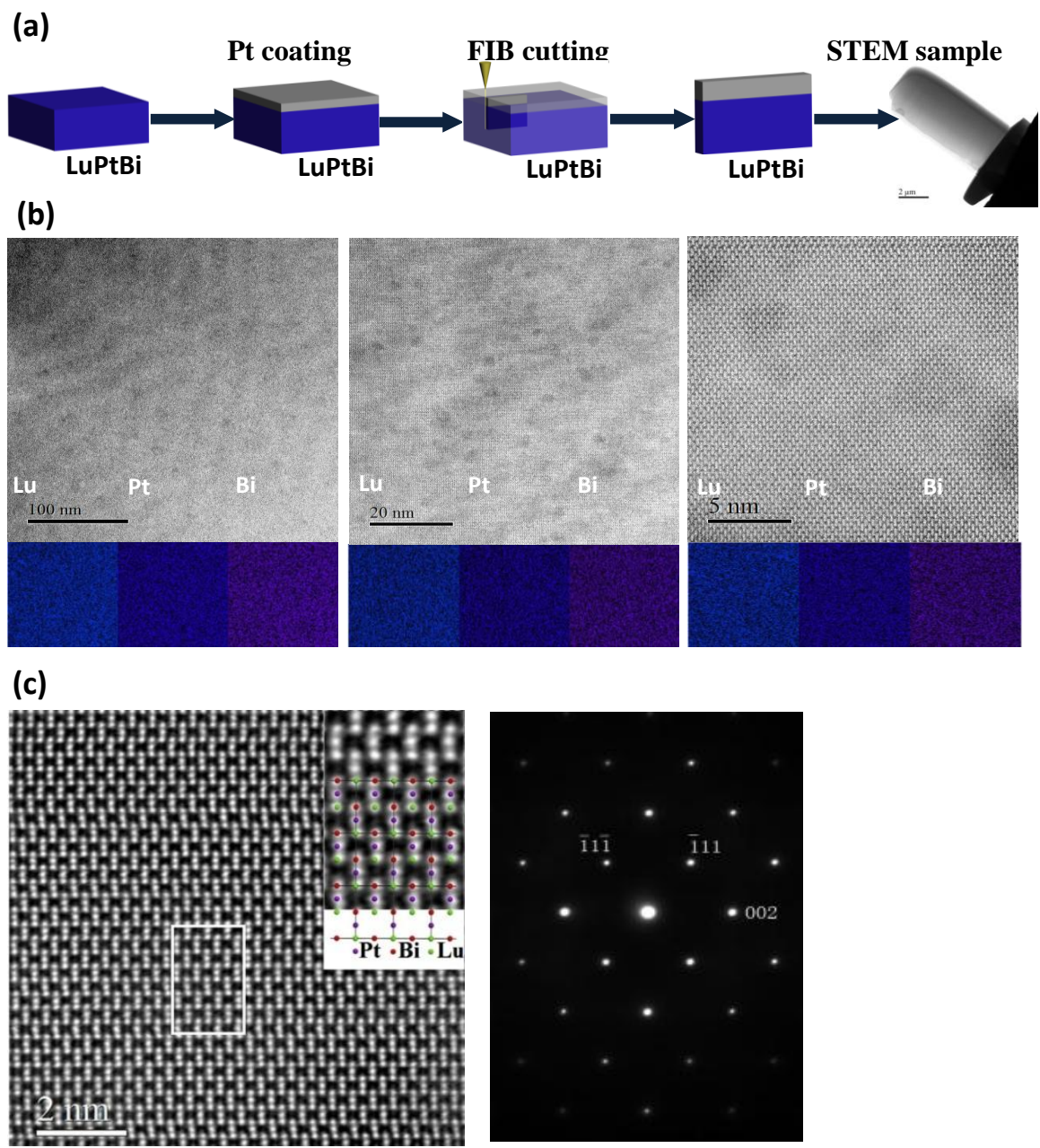

**FIG. 3.** (color online) (a) The schematic for the focused ion beam (FIB) process along the [110] zone axis; (b) Top plane: STEM images at different scales along the [100] stacking axis, where no evidence of Bi thin layers or nanoclusters were observed; Bottom plane: Elemental maps of Lu, Pt and Bi acquired by scanning EDX, demonstrating the uniform distribution of Lu, Pt and Bi across the surface. (c) The atomic-resolution STEM image in along the [100] stacking axis and selected diffraction patterns from the same plate, confirming the absence of Bi impurities.

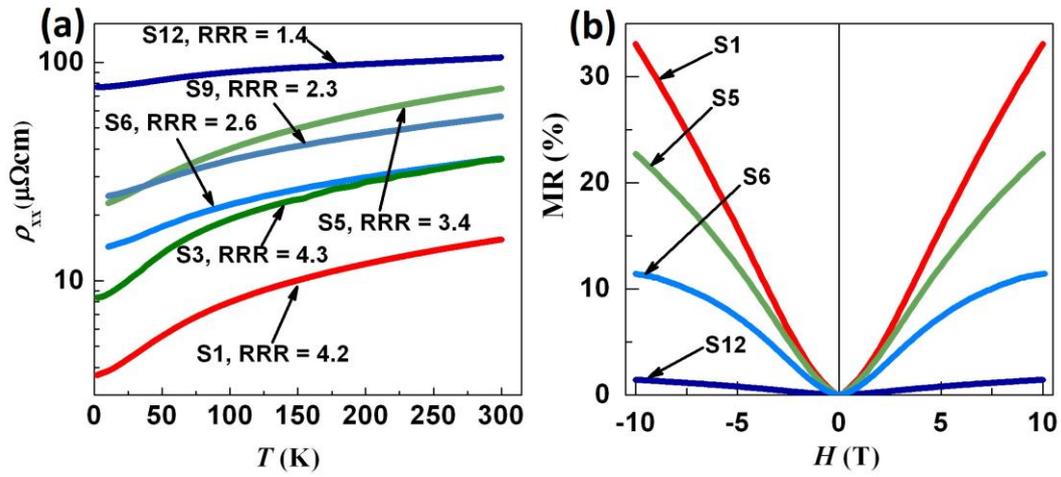

**FIG. 4.** (color online) (a) Temperature dependence of zero-field resistivity $\rho_{xx}$ curves for six representative crystals with various RRRs. (b) MR curves for samples S1, S5, S6 and S12 measured at 2 K.

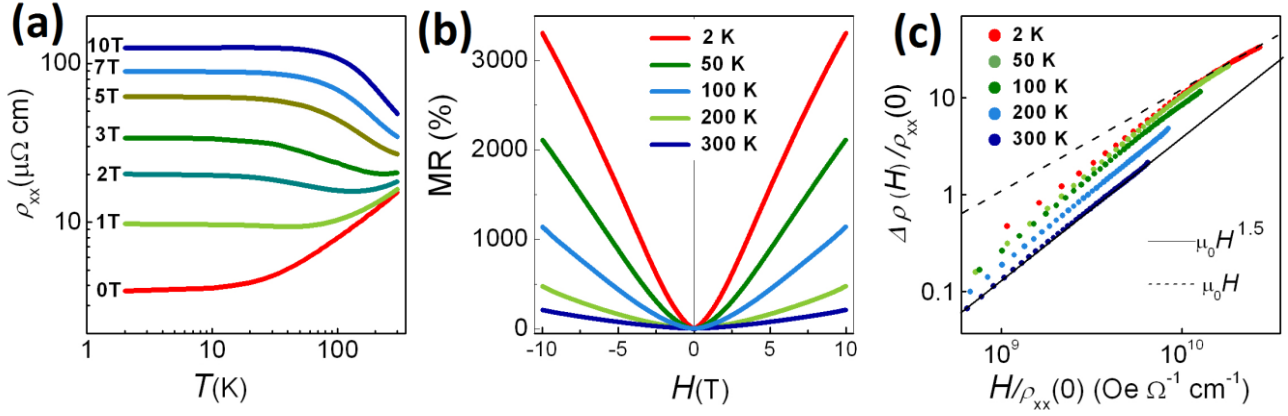

**FIG. 5.** (color online) (a) Temperature dependence of the resistivity ($\rho_{xx}$) of sample S1 at magnetic fields ranging from 0 to 10 T. At 0 T, $\rho_{xx}$ exhibits a metallic behavior. The external magnetic field increases the resistivity and changes the temperature-dependent behavior. (b), The magnetic field dependence of $\rho_{xx}$ at different temperatures. The MR ratio (MR=[$\rho_{xx}(H)$- $\rho_{xx}(0)$]/ $\rho_{xx}(0)$) increases monotonically with the increase of external magnetic field, without a saturation up to 10 T. A high MR ratio of 3200% and 260% is obtained at 2 and 300 K, respectively. (c), The Kohler plot: the MR ratio as a function of $H/\rho_{xx}(0)$. The solid and dashed lines are the fitted lines with the equations MR= $aH^{1.5}$ and b $H$, respectively. The MR data can be scaled by the equation MR= $aH^{1.5}$ at high temperatures and very low-field regime at low temperatures below 50 K but completely deviates from the Kohler's rule. At low temperatures and high-field regime, all the MR data collapse onto a single universal curve scaled linearly with $H$.

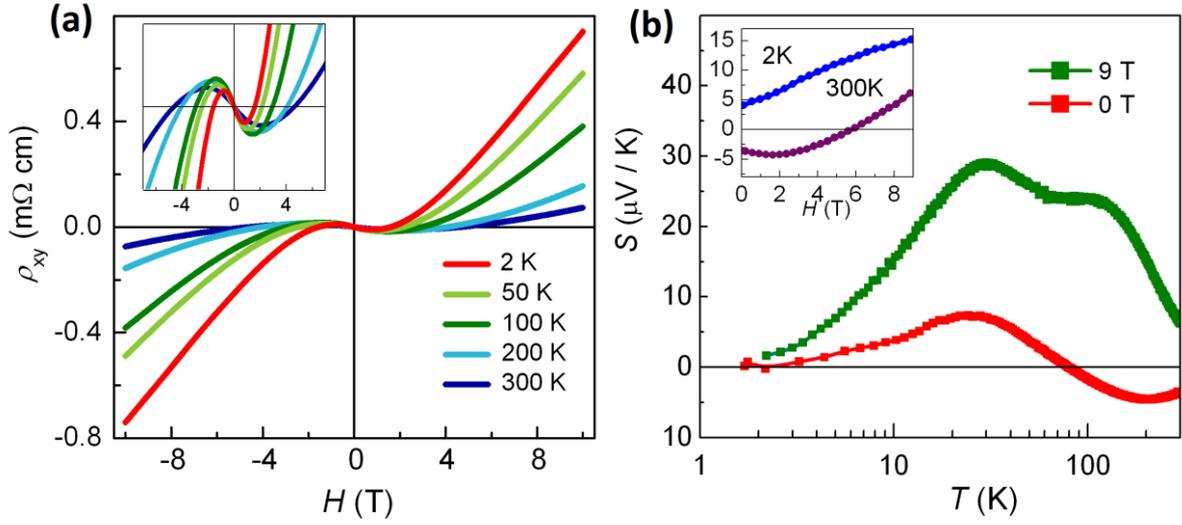

**FIG. 6.** (color online) (a) The magnetic field dependence of Hall resistivity ($\rho_{xy}$) for sampe S1 at temperatures ranging from 300 to 2 K. The inset presents the amplification of the region where $\rho_{xy}$ is close to zero and clearly exhibits the sign changing in $\rho_{xy}$. At 300 K, the negative slope of $\rho_{xy}$ indicates the predominance of electron carrier under a magnetic field below 2 T. However, the sign changes to positive value at higher magnetic field, indicating the further control of hole carrier. (**b**), The temperature dependence of Seebeck coefficient (*S*) at 0 and 9 T. The inset exhibits the magnetic field dependence of *S* at 10 and 300 K. At 0 T, the Seebeck coefficient changes the sign near 80 K, confirming the two-carrier model. On the other hand, the Seebeck coefficient only shows the positive value under a magnetic field of 9 T. This indicates the hole carrier dominates at high fields, agreeing well with Hall resistivity results.

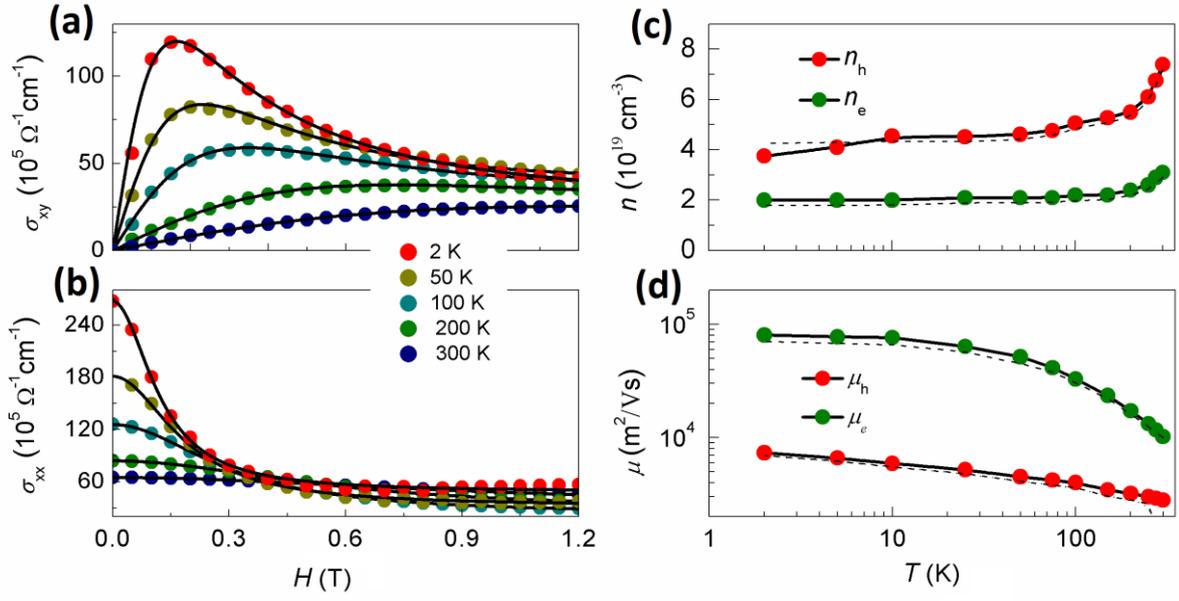

**FIG. 7.** (color online) [(a),(b)] Magnetic-field dependence of Hall conductivity $\sigma_{xy}$ and longitudinal conductivity $\sigma_{xx}$ for sample S1. The solid curves are the results of calculations using the two-carrier model (see the Supplemental Material [30]). (c) Temperature dependence of electron concentration $n_e$ and hole concentration $n_h$ estimated from conductivity $\sigma_{xy}$. (d) Temperature dependence of electron mobility $\mu_e$ and hole mobility $\mu_h$ estimated from conductivity $\sigma_{xy}$.

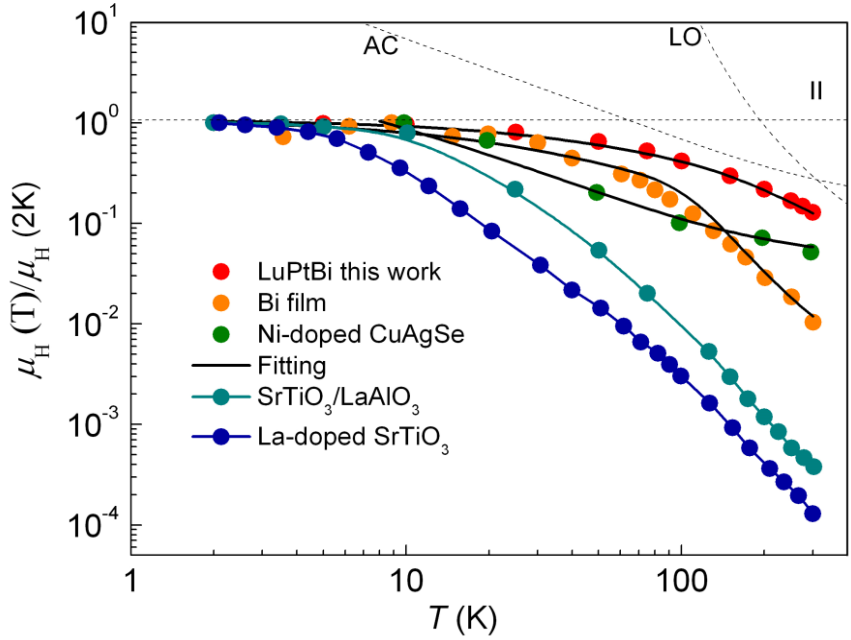

**FIG. 8.** (color online) The normalized $\mu_H(T)/\mu_H(2\ \text{K})$ of LuPtBi is compared with several of the most researched high-mobility systems. Clearly, the mobility of the SrTiO$_3$/LaAlO$_3$ interface (blue) and La-doped SrTiO$_3$ films (dark cyan) dropped rapidly with increasing temperature, while for semimetal systems, such as Bi film (orange), Ni-doped CuAgSe (green), and LuPtBi (red), temperature dependence is weak, preserving relatively high carrier mobility at room temperature. The solid lines are the fitting curves (see text for more details).

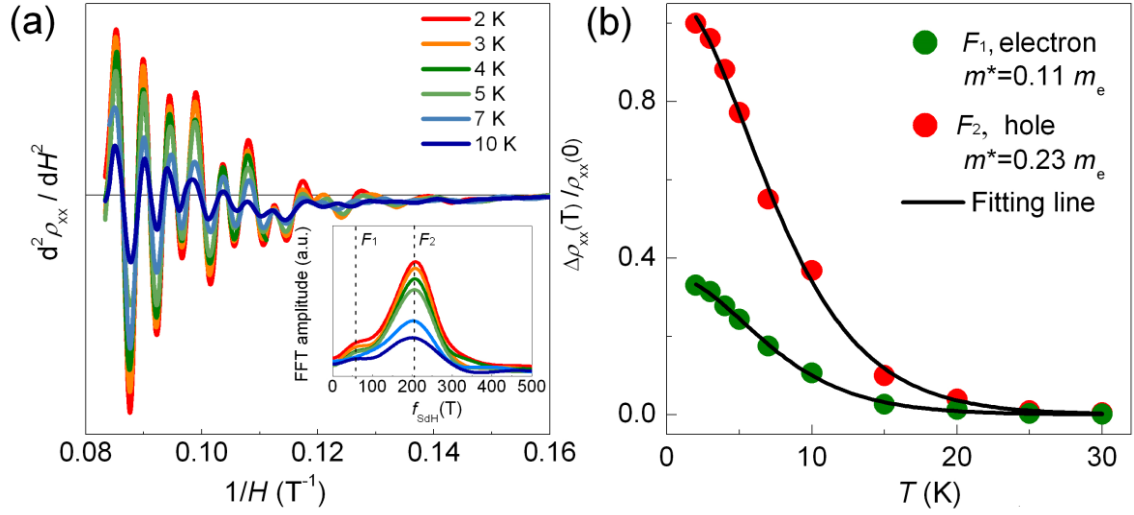

**FIG. 9.** (color online) (a) Temperature dependence of the SdH oscillations $d^2\rho_{xx}/dH^2$ as a function of the inverse magnetic field $1/H$. The inset shows the fast Fourier transform (FFT) spectra of SdH oscillations at different temperatures, in which $F_1$ and $F_2$ denote the electron and hole Fermi pockets, respectively. (b) Temperature dependence of the normalized FFT amplitude for the electron and hole Fermi pockets. The red and blue solid lines are the fits with the standard Lifshitz-Kosevich formula, yielding the small effective masses $0.11\,m_e$ and $0.23\,m_e$ for the electron and hole Fermi pockets, respectively.

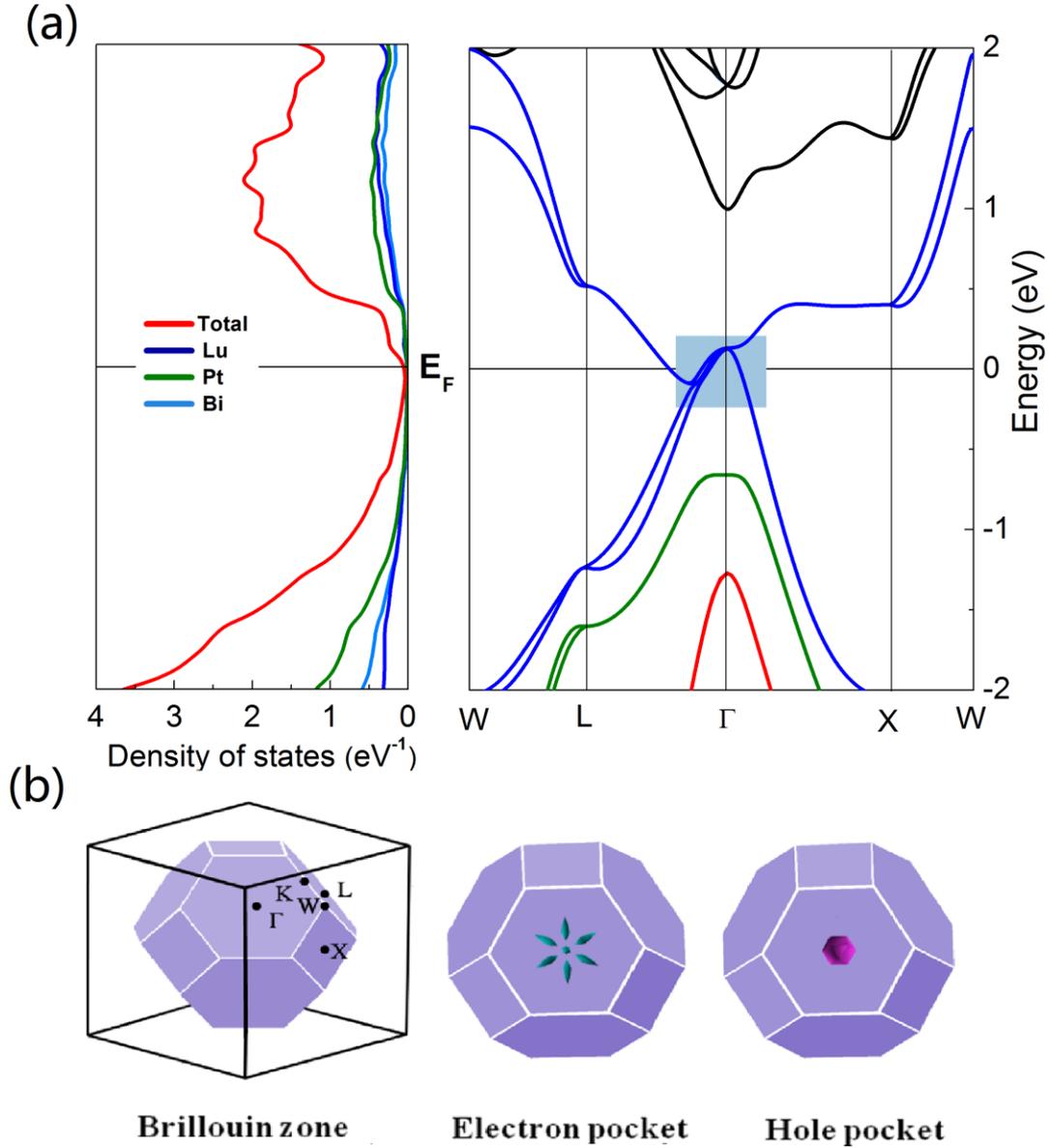

**FIG. 10.** (color online) (a) The total density of states (DOS) and local DOS from Lu, Pt, and Bi in the LuPtBi complex. The main peaks of total DOS are located far from the Fermi level and Lu, Pt and Bi contribute equally to the total DOS. (b) The calculated band structure of LuPtBi. The black line at E = 0 shows Fermi level position. The coexistence of electron and hole pockets agrees well with the two-carrier analysis. (c) The bulk Brillouin zone (left plane) and the Fermi surface of bulk LuPtBi showing electron (middle plane) and hole (right plane) pockets.

# Supplemental Materials

High Electron Mobility and Large Magnetoresistance in the Half-Heulser Semimetal LuPtBi

Zhipeng Hou[1,2], Wenhong Wang[1,*], Guizhou Xu[1], Xiaoming Zhang[1], Zhiyang Wei[1], Shipeng Shen[1], Enke Liu[1], Yuan Yao[1], Yisheng Chai[1], Young Sun[1], Xuekui Xi[1], Wenquan Wang[2], Zhongyuan Liu[3], Guangheng Wu[1] and Xi-xiang Zhang[4]

[1]State Key Laboratory for Magnetism, Beijing National Laboratory for Condensed Matter Physics, Institute of Physics, Chinese Academy of Sciences, Beijing 100190, China

[2]College of Physics, Jilin University, Changchun 130023, China

[3]State Key Laboratory of Metastable Material Sciences and Technology, Yanshan University, Qinhuangdao 066004, China

[4]Physical Science and Engineering, King Abdullah University of Science and Technology (KAUST), Thuwal 23955-6900, Saudi Arabia.

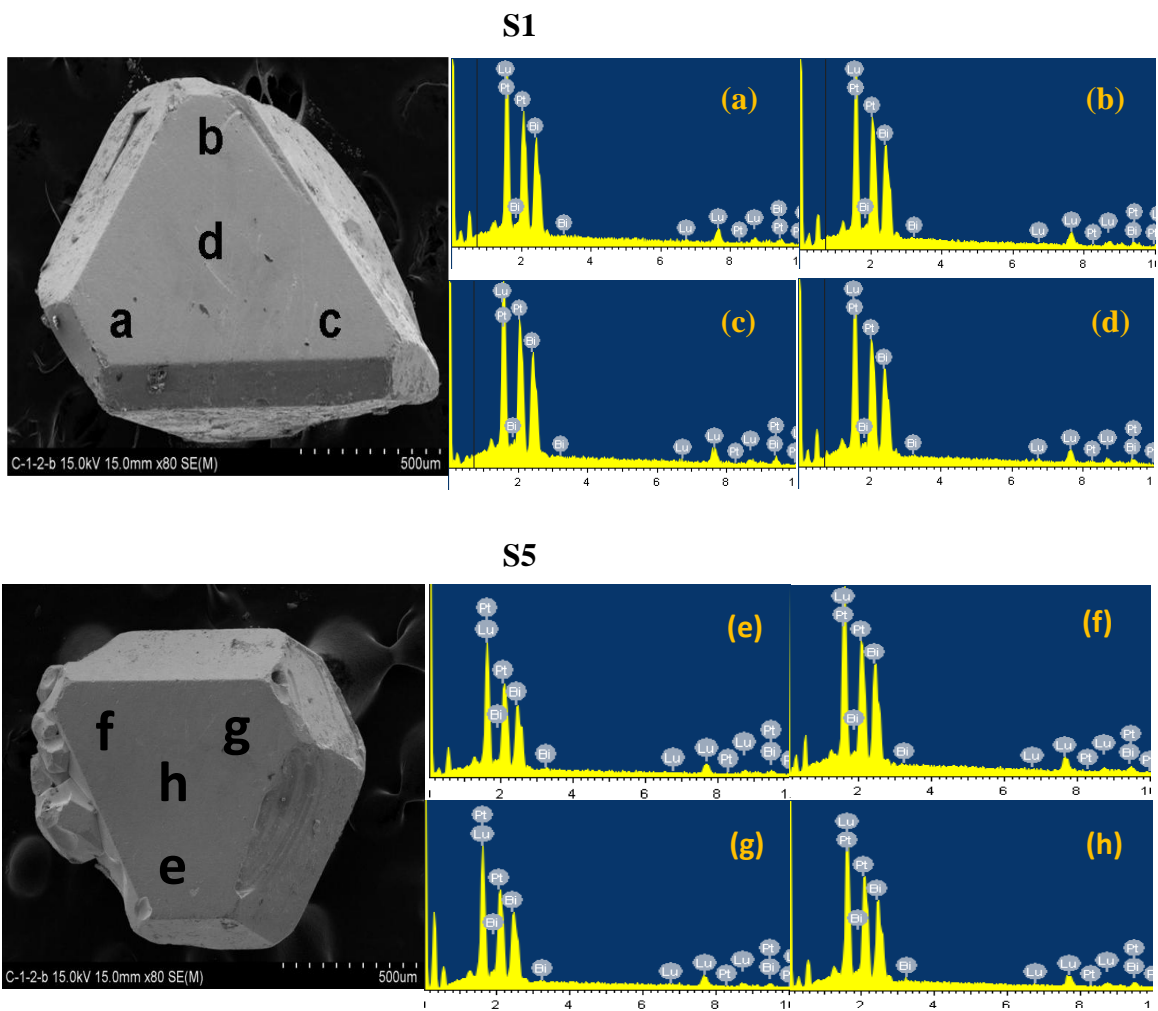

**FIG. S1.** EDX spectras of samples S1 (a, b, c, d) and S5 (e, f, g, h). EDX analyses were performed at numbered points on the crystal surface and the corresponding results are listed in Table S1. The average atom percentages of Lu: Pt: Bi were 31.96%: 32.94% : 35.10% and 31.94%: 33.43%: 34.63% for S1 and S5, respectively. Notably, EDX results are only semi-quantitative and a measurement error of 1- 2% should be considered.

| | S1 | | |
|---|---|---|---|
| Position | Lu (at.%) | Pt (at.%) | Bi (at.%) |
| a | 32.16 | 32.59 | 35.25 |
| b | 31.59 | 32.85 | 35.56 |
| c | 31.36 | 33.92 | 34.73 |
| d | 32.75 | 32.38 | 34.87 |
| Average | 31.96 | 32.94 | 35.10 |
| | S5 | | |
| Position | Lu (at.%) | Pt (at.%) | Bi (at.%) |
| a | 31.86 | 33.69 | 34.45 |
| b | 31.83 | 33.59 | 34.58 |
| c | 31.88 | 33.38 | 34.74 |
| d | 31.77 | 33.45 | 34.78 |
| Average | 31.84 | 33.52 | 34.69 |

**Table S1.** Chemical compositions of S1 and S5 samples determined by EDX at different positions of crystal surfaces.

| Sample | t mm | $\rho_{xx}$ μΩcm | RRR $R_{300K}/R_{2K}$ | MR 2K, 10T | $\mu_e$ cm$^2$V$^{-1}$s$^{-1}$ | $\mu_h$ cm$^2$V$^{-1}$s$^{-1}$ | $n_e$ 10$^{19}$cm$^{-3}$ | $n_h$ 10$^{19}$cm$^{-3}$ |
|---|---|---|---|---|---|---|---|---|
| S1 | 0.21 | 3.70 | 4.3 | 3200% | 79000 | 7320 | 2.0 | 3.7 |
| S3 | 0.23 | 8.36 | 4.1 | 3000% | 74000 | 5400 | 1.7 | 3.0 |
| S5 | 0.19 | 22.7 | 3.4 | 1900% | 31000 | 5000 | 1.2 | 2.5 |
| S6 | 0.23 | 14.3 | 2.6 | 1000% | 12000 | 4300 | 1.0 | 2.3 |
| S9 | 0.21 | 24.5 | 2.3 | 353% | 8200 | 2500 | 0.4 | 2.0 |
| S12 | 0.25 | 76.8 | 1.4 | 136% | - | 2400 | - | 3.0 |

**Table S2**.  *t* is the thickness of the sample. $\rho_{xx}$ is the resistivity at 2K. *RRR* represents the ratio of $\rho_{xx}(300K)/\rho_{xx}(2K)$. The mobilities ($\mu_e$ and $\mu_h$) and carrier concentrations ($n_e$ and $n_h$) of S1, S3, S5, S6, S9 and S12 were established by two-carrier model but one-band model for S12 sample.

| S2 | | | | | |
|---|---|---|---|---|---|
| t (mm) | $\rho_{xx}$ ($\mu\Omega$cm) | RRR ($R_{300K}/R_{2K}$) | MR (2K, 10T) | $\mu_e$ (cm$^2$V$^{-1}$s$^{-1}$) | $\mu_h$ (cm$^2$V$^{-1}$s$^{-1}$) |
| 0.60 | 7.82 | 3.8 | 2800% | 65000 | 5500 |
| 0.35 | 8.36 | 3.77 | 2650% | 63500 | 5400 |
| 0.15 | 8.45 | 3.75 | 2600% | 63000 | 5400 |
| S6 | | | | | |
| t (mm) | $\rho_{xx}$ ($\mu\Omega$cm) | RRR ($R_{300K}/R_{2K}$) | MR (2K, 10T) | $\mu_e$ (cm$^2$V$^{-1}$s$^{-1}$) | $\mu_h$ (cm$^2$V$^{-1}$s$^{-1}$) |
| 0.62 | 21.4 | 2.6 | 670% | 11000 | 3800 |
| 0.43 | 22.5 | 2.56 | 640% | 10800 | 3500 |
| 0.21 | 21.7 | 2.5 | 645% | 10870 | 3450 |

**Table S3**. Thickness dependence of the transport parameters in samples S2 and S6.

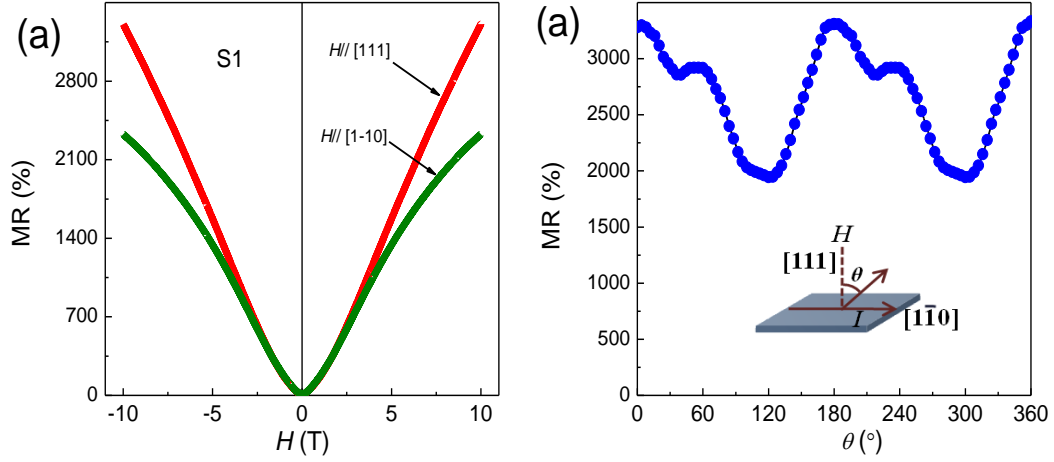

**FIG. S2.** (a) The MR ratio for Sample S1 at different orientations. The $\Delta\rho_{xx}(H)/\rho_{xx}(0)$ under a perpendicular magnetic field (*H* perpendicular to *I* and ab-plane) is shown by the red line, and $\Delta\rho_{xx}(H)/\rho_{xx}(0)$ at 2 K under a longitudinal magnetic field (*H* perpendicular to *I* and parallel to the ab-plane) is shown by the black line. At 2 K and under a magnetic field of 10 T, a transverse MR ratio of 3200% was obtained, while the longitudinal MR decreased to 2230%, indicating that the angle between the magnetic field and the [111] direction can affect the MR ratio. (b) Observed azimuthal field-angle dependence of the MR ratio of LuPtBi at 2 K and 10 T. Here, magnetic field *H* is rotated from the [111] to the [1-10] direction, as shown by the inset.

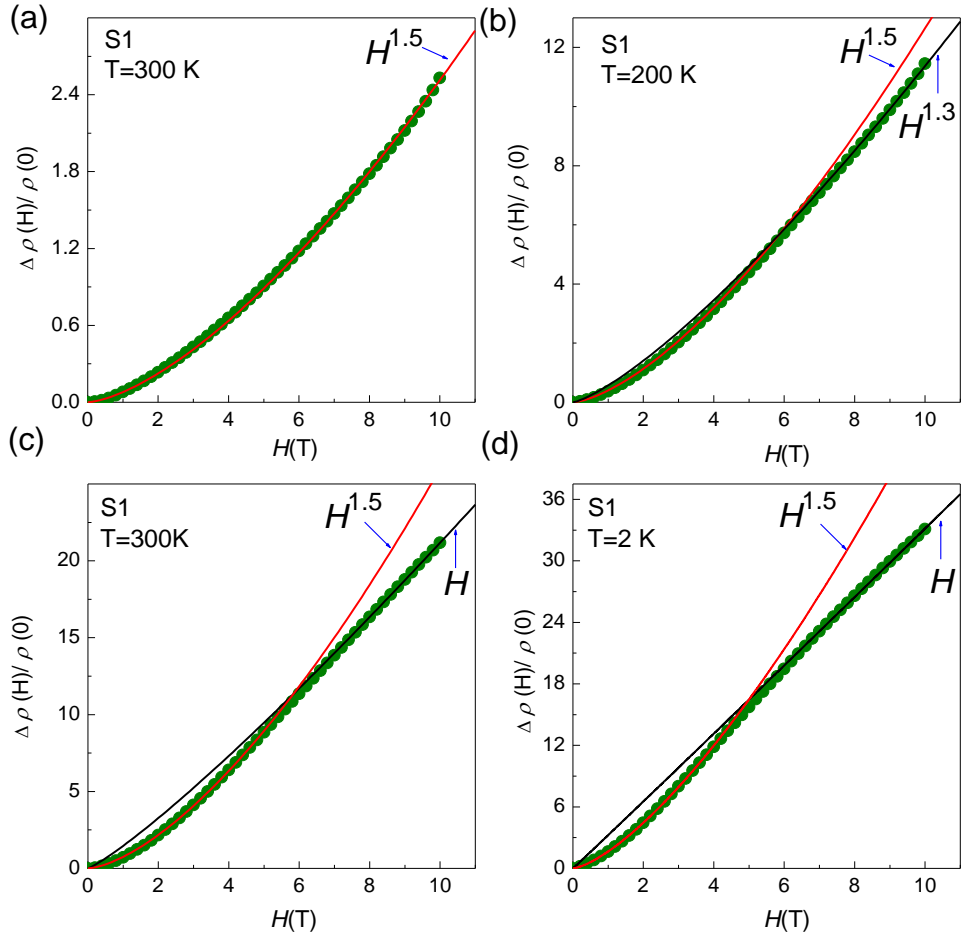

**FIG. S3.** The magnetic field dependence of the MR ratio $\Delta\rho_{xx}(H)/\rho_{xx}(0)$ for LuPtBi single crystals at a series of temperatures, $T$ = (a) 300 K, (b) 100 K, (c) 50 K and (d) 2 K. Green circles are the experimental data and the solid lines (red and black lines) are fit data fit with MR = $aH^b$. At 300K, MR can be fitted well with $aH^{1.5}$; however, MR at a high-field regime gradually deviates from $aH^{1.5}$ with decreasing temperature and shows a linear MR behavior at low temperatures (between 50 K and 2 K) and high-field regimes.

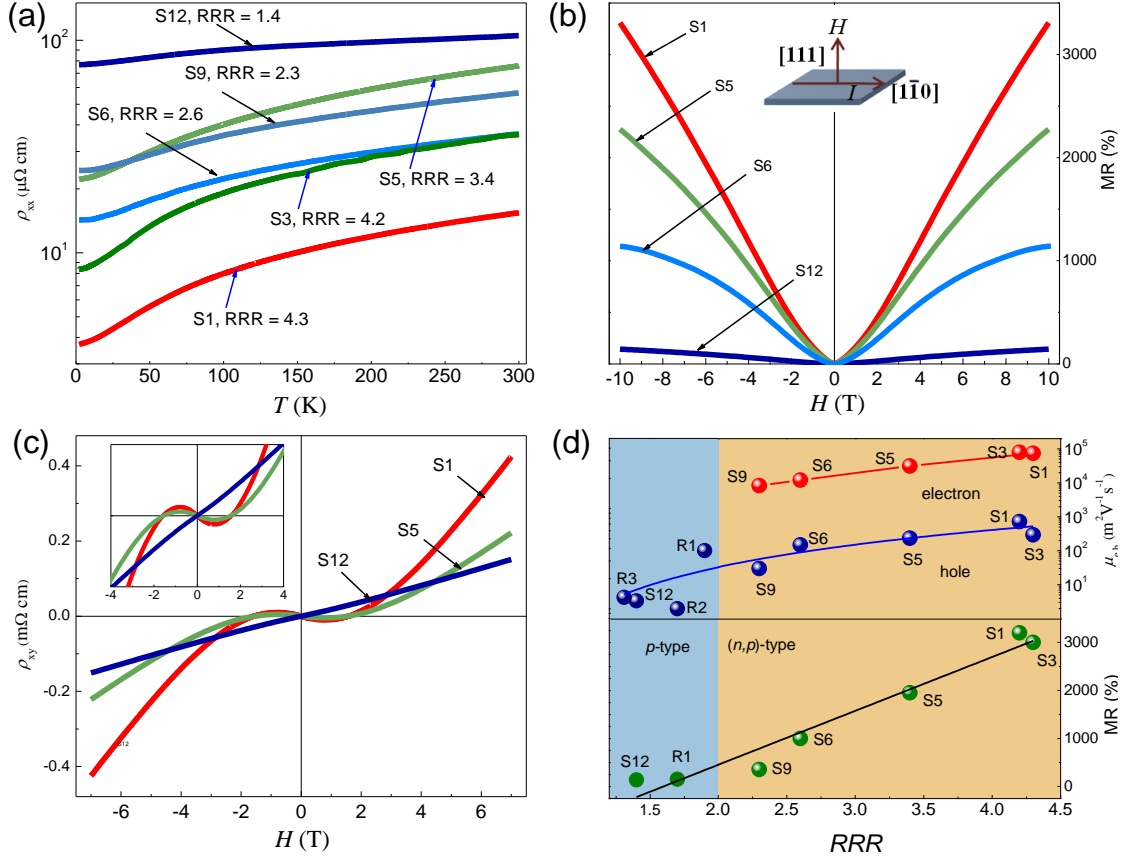

**FIG. S4**. (a) Temperature dependence of zero-field resistivity $\rho_{xx}$ curves for six representative crystals with various RRRs. (b) MR curves for samples S1, S5, S6 and S12 measured at 2 K. (c) Hall resistivity $\rho_{xy}$ curves for samples S1, S3, and S12 measured at 2 K. The upper inset shows an expanded view of the low $H$ region, and the lower inset is a high-resolution STEM image for sample S3. (d) Upper plane: electron (red solid) and hole (blue solid) mobilities versus RRR. Lower plane: MR values (green solid) versus RRR. Orange represents the coexistence of electron and hole carriers, whereas the blue region indicates a hole-dominated transport region.

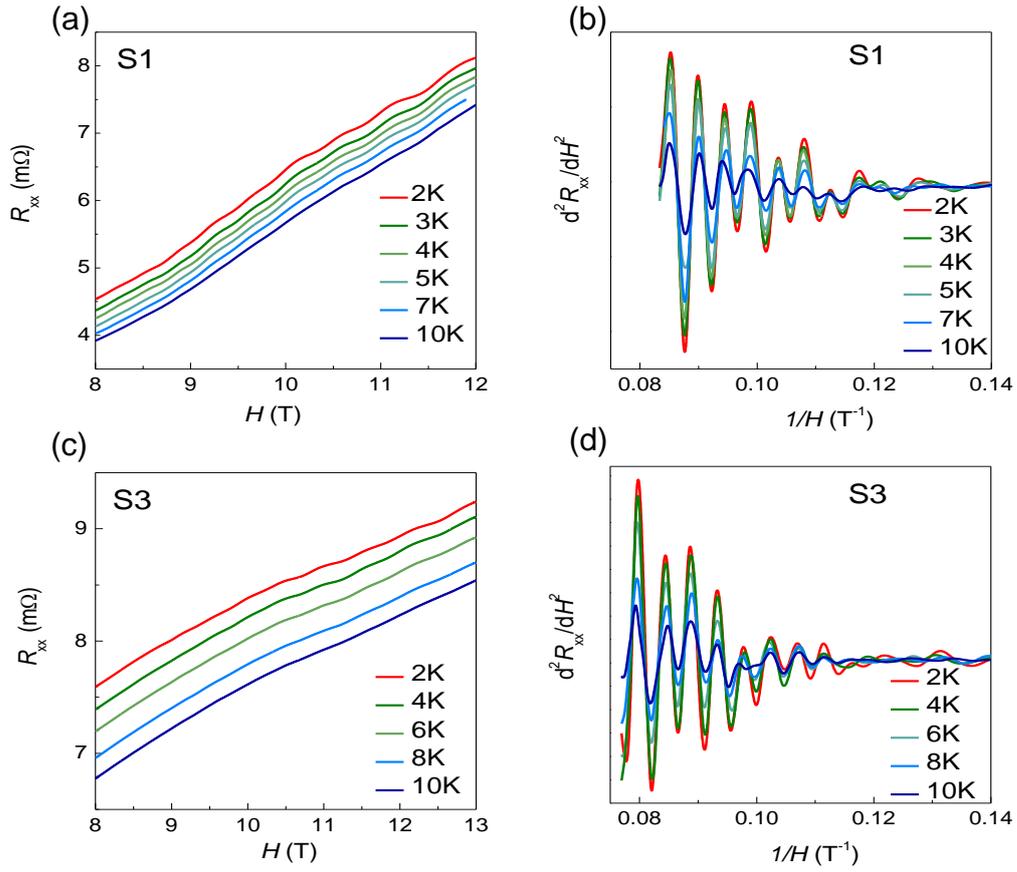

**FIG. S5.** The temperature dependence of SdH quantum oscillations at high magnetic fields for samples S1 and S3. The field is applied perpendicular to the [111] direction. At high fields, as shown in (a) and (b), the clear SdH oscillations are superimposed on a huge background of positive MR. The oscillations become apparent in the second derivative $d^2R_{xx}/dH^2$ as a function of the inverse magnetic field $1/H$ as shown in (c) and (d).

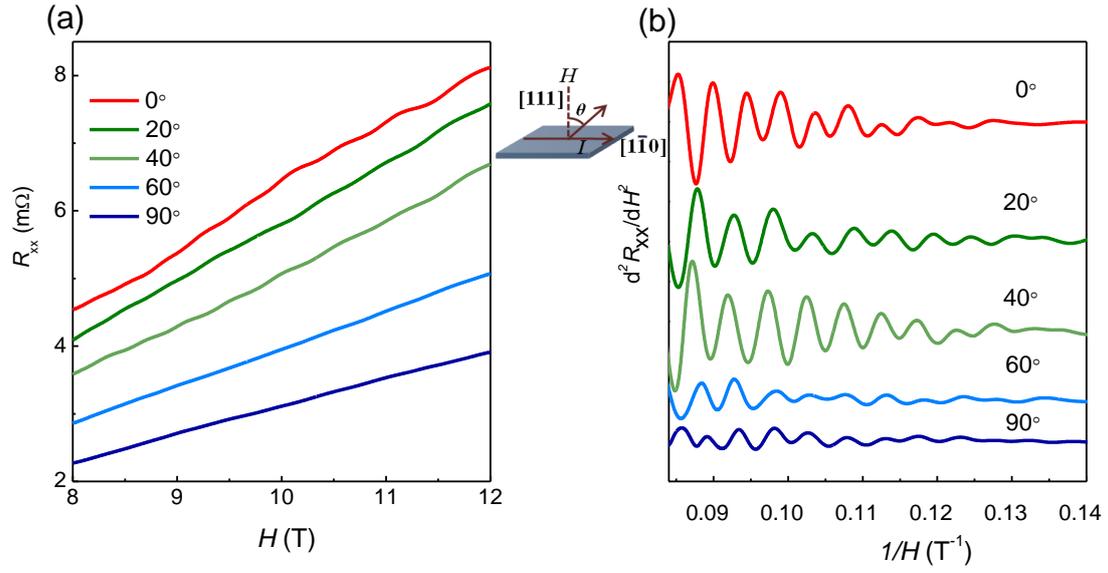

**FIG. S6.** (a) Angle dependence of resistance of sample S1 at high fields. The magnetic field $H$ is rotated in the z-y plane as shown in the inset. When we rotate the angle $\theta$ from the [111] to the [1-10] direction, as shown in (b), distinct SdH oscillations are observed in the second derivative $d^2R_{xx}/dH^2$ as a function of the inverse magnetic field $1/H$.

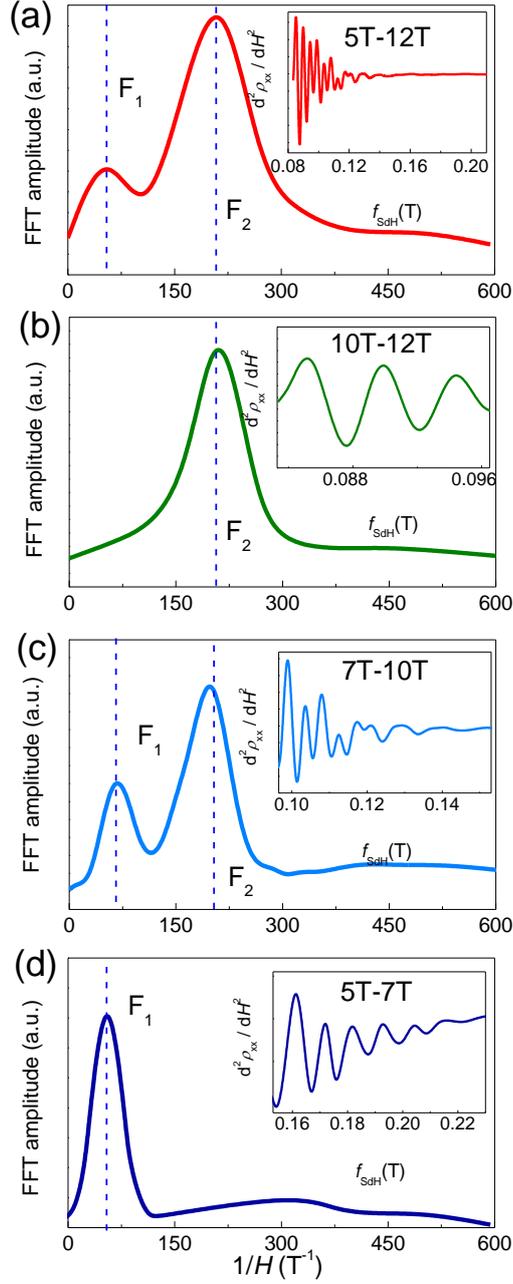

**FIG. S7.** (a) FFT in the magnetic field range of 5T-12T, two peaks at $F_1=80$ T and $F_2=200$ T are found. (b) FFT in the magnetic field range of 10T-12T, only one peak at $F_2=200$ T is found. Since the holes carriers dominate the transport properties at high fields, the F2 peak corresponds to the SdH oscillations of hole. (c) FFT in the magnetic field range of 7T-10T, $F_1$ at 80T together with F2 peak appear. (c) FFT in the magnetic field range of 5T-7T, only $F_1$ peak can be found. Since the electron exhibits higher mobility, it should show SdH oscillations at lower magnetic field. Therefore, the F1 peak corresponds to SdH oscillations of electron.